\documentclass[10pt, conference]{IEEEtran}
\usepackage{cite}
\usepackage{amsmath,amssymb,amsfonts}
\usepackage{graphicx}
\usepackage{textcomp}
\usepackage{xcolor}
\usepackage{booktabs}
\usepackage{algorithm}
\usepackage{algorithmic}
\usepackage{mathtools}

\begin{document}

\title{Dynamic Authorization for Knowledge-Base Agents in 6G}

\author{
\IEEEauthorblockN{Loay Abdelrazek\IEEEauthorrefmark{1}, Leyli Kara\c{c}ay\IEEEauthorrefmark{2}, Marin Orli\'{c}\IEEEauthorrefmark{1}}
\IEEEauthorblockA{\IEEEauthorrefmark{1}Ericsson, Sweden; \IEEEauthorrefmark{2}Ericsson, T\"{u}rkiye \\ 
Email: \{first.last\}@ericsson.com}
}

\maketitle

\begin{abstract}
As 6G architectures transition toward decentralized Multi-Agent Systems (MAS), ensuring secure access to shared Knowledge Bases (KB) is critical. Traditional authorization models like RBAC fail to provide the granularity required for autonomous agents interacting with Semantic-based data. This work proposes a hybrid authorization framework that integrates roles and First-Order Logic (FOL) predicates to enforce zero-trust principles at the knowledge-graph level. We eliminate permission inheritance by enforcing authorization at the triple level (Subject-Predicate-Object), ensuring agents only access metadata required for their specific functional lifecycle.
\end{abstract}


\section{Introduction}
The evolution of 6G necessitates a shift toward open, multi-vendor autonomous systems where decentralized agents collaborate via a shared Knowledge Base (KB). These agents utilize Knowledge Graphs (KG) to store network state, ontologies, and operational metadata. However, existing Role-Based Access Control (RBAC) models suffer from "permission inheritance," where an agent assigned a general role inherits access to all predicates associated with that role. 

In a multi-vendor 6G environment, a compromised agent could exploit this inheritance to perform lateral movement across the KB. We address this by proposing a dynamic, triple-level authorization model that aligns with NIST Zero Trust Architecture (ZTA) standards \cite{b1}, ensuring "least privilege" access for every agent query.

\section{System Model and Representation}
\subsection{Multi-Agent Knowledge Interaction}
Our framework assumes a 6G environment where specialized agents (e.g., grounders, monitors, optimizers) interact with an Intent Management Function (IMF) or a broader Knowledge Management system. The KB represents facts as triples $(s, p, o)$. To facilitate granular authorization, metadata is categorized into two types: 1) \textbf{Static Knowledge} (deployment-time ontologies and agent definitions) and 2) \textbf{Dynamic Knowledge} (runtime attributes such as session-specific resource identifiers).
\section{Proposed Hybrid Authorization Framework}
The core contribution is a hybrid model that shifts the Policy Decision Point (PDP) directly into the logical reasoning layer.

\subsection{Formal Logic-Based Authorization}
We define authorization through First-Order Logic (FOL) predicates. Access to a specific triple is granted only if the predicate $p$ matches the agent's dynamic security profile $G_{\text{auth}}$:
{\small
\begin{equation}
\text{Auth}(A, s, p, o) \leftarrow \text{Req}(A, p) \land \text{Profile}(A, p) \land \text{Role}(A, R)
\end{equation}}

This logic ensures that authorization is not just based on "who" the agent is (Role), but "what" specific relationship it is attempting to access (Predicate). A comparison with existing models can be found in Table \ref{tab:comparison} below. Additionally, to prevent ontology ``crawling'' we enforce:
{\small
\begin{equation}
\forall Q \in \text{AgentQueries}: \text{isVariable}(p) \Rightarrow \text{AccessDenied}
\end{equation}}
By restricting variable predicates, we ensure agents cannot discover hidden metadata relationships outside their authorized functional scope.

\begin{table}[htbp]
\centering
\vspace{-2mm}
\caption{Comparison of Authorization Models for 6G}
\begin{center}
\begin{tabular}{|l|c|c|c|}
\hline
\textbf{Feature} & \textbf{RBAC} & \textbf{RelBAC} & \textbf{Proposed Hybrid} \\ \hline
Granularity & Role-level & Relationship & \textbf{Triple-level} \\ \hline
Inheritance & Yes & Yes & \textbf{No (Denied)} \\ \hline
Context-Aware & Low & Medium & \textbf{High (FOL-based)} \\ \hline
6G Readiness & Low & Partial & \textbf{Full (Dynamic)} \\ \hline
\end{tabular}
\end{center}
\label{tab:comparison}
\vspace{-2mm}
\end{table}

\section{Continuous Enforcement Mechanism}
\begin{figure*}[t]
\centering
\includegraphics[width=0.7\textwidth]{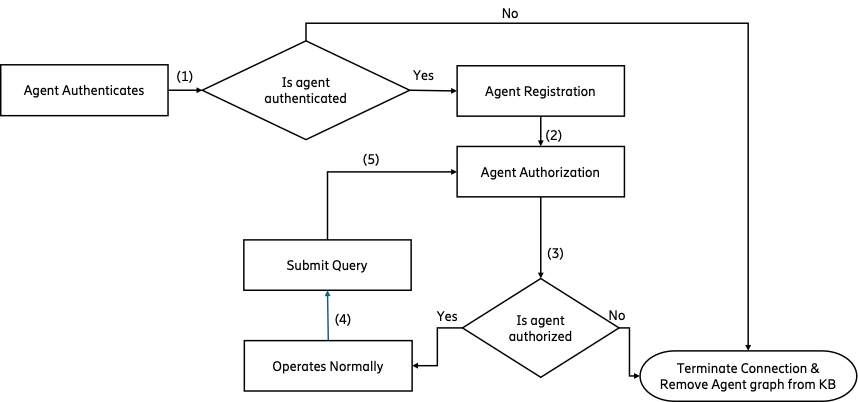}
\vspace{-2mm}
\caption{The Authorization Lifecycle: Registration to Query Enforcement.}
\label{fig:flow}
\vspace{-4mm}
\end{figure*}

The framework shifts the security boundary from the network perimeter to the individual data triple. The Policy Enforcement Point (PEP) is natively integrated into the logical reasoner, treating every agent interaction as a unique, time-bound session that requires continuous validation.
Authorization occurs at two critical stages: during agent registration and upon every subsequent query. The authorization lifecyle is illustrated in Figure \ref{fig:flow}.

\subsection{Query Pattern Extraction and Deconstruction}
When a Knowledge-Base agent $A$ issues a request (typically via SPARQL or a logic-based query language), the system first deconstructs the query $Q$ into its constituent triple patterns:
{\small
\begin{equation}
P_{req} = \{ (s_1, p_1, o_1), (s_2, p_2, o_2), \dots, (s_n, p_n, o_n) \}
\end{equation}}
The extraction process identifies which predicates ($p_i$) are being accessed. This is the critical juncture for \textbf{Wildcard Protection}: if the query uses a variable for a predicate (e.g., \textit{?p}), the reasoner immediately flags the request as a potential "crawling" attempt and denies access, unless the agent is operating within a specific administrative ontology scope.

\subsection{Logical Intersection and Inference Check}
The core of the enforcement logic is the mathematical intersection between the requested patterns $P_{req}$ and the agent’s dynamically generated authorization profile $G_{auth}$. The Reasoner performs an inference check to verify that for every triple in the query, a corresponding authorization assertion exists:
{\small
\begin{equation}
\forall (s, p, o) \in P_{req} : G_{auth} \cup \mathcal{O} \vdash \text{CanAccess}(A, p)
\end{equation}}
where $\mathcal{O}$ represents the global security ontology. Unlike static filters, this inference-based check allows the system to handle complex relationships, such as allowing access to a predicate only if it is a sub-property of a previously authorized functional class.

\subsection{Semantic Result Refinement}
Even after query pattern validation, the Reasoner performs a final enforcement layer by refining the result set. By analyzing the semantic context of the returned triples—such as their association with specific Intent instances or Domain IDs—the framework prunes data that falls outside the agent's task-specific boundaries. This ensures that Agent A, while authorized to access a general class of metadata, is restricted to its specific operational instance, preventing cross-tenant data leakage.

The complete procedural logic for this continuous verification and refinement is formalized in Algorithm 1.

\begin{algorithm}
\caption{Dynamic Agent Authorization Enforcement}
\begin{algorithmic}
\small
\STATE \textbf{Input:} Agent $A$, Query $Q$, Auth Profile $G_{\text{auth}}$
\STATE \textbf{Output:} Filtered Result Set $R$ or Access Denied
\STATE $P_{\text{req}} \leftarrow \text{ExtractTriplePatterns}(Q)$
\FORALL{$t_i \in P_{\text{req}}$}
    \IF{$t_i.\text{predicate} \text{ is variable}$}
        \RETURN \textbf{Access Denied}
    \ENDIF
    \IF{\textbf{not} InferenceCheck($A, t_i.\text{predicate}, G_{\text{auth}}$)}
        \STATE LogViolation($A, t_i.\text{predicate}$)
        \RETURN \textbf{Revoke Session}
    \ENDIF
\ENDFOR
\STATE $R_{\text{raw}} \leftarrow \text{ExecuteQuery}(Q, G_{\text{kb}})$
\STATE $R \leftarrow \text{ApplyContextualPruning}(R_{\text{raw}}, A.\text{context})$
\RETURN $R$
\end{algorithmic}
\end{algorithm}

\section{Security Analysis and Discussion}
The proposed hybrid framework effectively segments the multi-vendor Knowledge Base by treating every predicate as a distinct trust boundary. 

\subsection{Mitigation of Lateral Movement}
By eliminating permission inheritance, an attacker who compromises a "Monitor" agent is restricted only to the monitoring predicates registered in that specific agent's $G_{\text{auth}}$. They cannot pivot to access "Actuation" or "Grounding" triples, as those predicates lack the required FOL assertions in the active session profile.

\section{Dynamic Revocation}
In Zero Trust 6G, authorization is a transient permission revocable at the atomic level. Our framework implements a \textbf{Triggered Profile Invalidation} mechanism. If Algorithm 1 detects a violation—such as unauthorized access to actuation predicates—the Reasoner performs a real-time retraction of the agent's security assertions. 

The revocation state $\mathcal{R}$ is defined by the immediate removal of the agent's dynamic authorization graph $G_{\text{auth}}$ from active memory:
{\small
\begin{equation}
G_{\text{kb}} \coloneq G_{\text{kb}} \setminus G_{\text{auth}}^{(A)}
\end{equation}}
where $G_{\text{auth}}^{(A)}$ is the triple set derived during registration. Consequently, any subsequent query $Q$ fails the inference check as the required logical grounding is purged. This ``instant-kill'' capability prevents lateral movement in multi-vendor systems.

\section{Conclusion}
By integrating authorization into the reasoning layer, we provide a scalable, zero-trust mechanism for 6G KB agents. This model eliminates permission inheritance and provides the granularity required for secure multi-vendor autonomous operations.


\begin{thebibliography}{00}
\bibitem{b1} NIST, "Zero Trust Architecture," Special Publication 800-207, 2020.
\end{thebibliography}
\end{document}